\documentclass[a4paper,11pt]{article}
\usepackage{pos}
\usepackage{subcaption}

\title{Hyperon-Production Studies in $ p+p \rightarrow \Lambda + K^0_S + p + \pi^+ $ with HADES}
\ShortTitle{Hyperon-Production Studies in $ p+p \rightarrow \Lambda + K^0_S + p + \pi^+ $}

\author[a,b]{Snehankit Pattnaik}
\author[a]{Johan Messchendorp}
\author[a,b,c]{James Ritman}

\onbehalf{for the HADES collaboration}

\affiliation[a]{GSI Helmholtzzentrum für Schwerionenforschung GmbH,\\
  Planckstraße 1, Darmstadt, Germany}

\affiliation[b]{Fakultät für Physik und Astronomie, Ruhr-Universität Bochum,\\
Universitätsstraße 150, Bochum, Germany}

\affiliation[c]{Institut für Kernphysik (IKP), Forschungszentrum Jülich,\\
Wilhelm-Johnen-Straße, Jülich, Germany}

\emailAdd{s.pattnaik@gsi.de}
\emailAdd{j.messchendorp@gsi.de}
\emailAdd{j.ritman@gsi.de}

\abstract{
The production mechanism of hyperons in the proton-proton collisions is explored using exclusive process with the data sample collected by HADES. The clear sign for three intermediate resonances is observed using simulated samples. It opens possibility for further detailed research.
}

\FullConference{
}

\begin{document}
\maketitle

\section{Introduction} \label{intro} 
The non-perturbative nature of the strong interaction, {\it i.e.} quantum chromodynamics, challenges our understanding of the internal structure and interactions of baryons such as the proton and neutron. Experimental measurements of the baryon excitation spectrum and the couplings of baryons to various final states, including strangeness, allow us to probe their nature and to search for new baryon resonances. Hyperon production is of particular interest since it provides information about the role of $N^{*}$ resonances in strangeness production in nucleon-nucleon (NN) interactions. Therefore, it could be relevant to describe the dynamics of hyperon production in elementary (p+p) reactions as a reference for understanding many-body systems (such as dense, baryon-rich matter).

A recent partial wave analysis of data taken with HADES of the reaction $ p\text{(3.5 GeV)} +p \rightarrow p + K^+ + \Sigma^0$ provided new insights into various intermediate baryon resonances and their couplings to final states including hyperons \cite{A}. The analysis hints at the importance of a few baryon resonances coupling to the $\Sigma K$ final state, albeit with limited statistics.
This work focuses on the reaction $ p+p \rightarrow \Lambda + K^0_S + p + \pi^+$, analyzing the coupling of baryons to the $\Lambda K^0_S, \Lambda K(892), \Sigma(1385) K^0_S$ final states. This study will involve assessing contributions from isospin 3/2 to the final states through differential cross-section analyses, complementing earlier studies. 
Additionally, extracting a precise cross section value from these data also serves as a reference estimate as a background channel for the upper limit cross section estimations for cascade production of $\Xi^-$ at HADES.

\section{Experiment} \label{experiment}

\begin{figure}[h]
\centering
\includegraphics[width=0.75\textwidth]{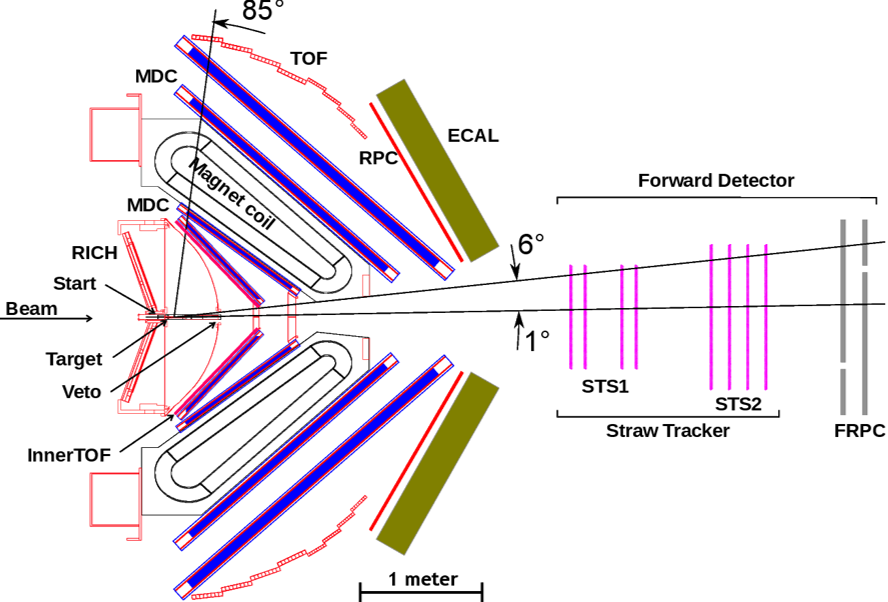}
\caption{Cross sectional view of the HADES spectrometer with the forward straw tube detector.}
\label{fig:had}
\end{figure}

The experiment discussed in this analysis is conducted with proton beams delivered with SIS18, with a kinetic beam energy of 4.5 GeV, at HADES in the GSI Helmholtz Center for Heavy Ion Research in Darmstadt, Germany \cite{B}. The HADES setup comprises six identical sectors, defined by superconducting coils, generating a toroidal magnetic field. This spectrometer covers an azimuthal acceptance of 85\% and polar angles from $\theta = 18^\circ$ to $\theta = 85^\circ$ as shown in Fig. \ref{fig:had}.

HADES consists of a combination of detectors, starting with two sets of multi-wire drift chambers (MDC) placed upstream and downstream of the magnetic field. These chambers enable precise momentum reconstruction and provide particle identification through energy loss measurements. Two time-of-flight detectors (ToF) and resistive plate chambers (RPC), positioned in different angular regions, improve stop-time measurements. With the addition of forward straw tube tracking, HADES is well-suited for hadronic studies in the SIS18 energy regime particularly for hyperon measurements. Additionally, the Ring Imaging Cherenkov (RICH) detector provides excellent di-lepton selection in a field-free region. Lastly, an electromagnetic calorimeter is used for gamma reconstruction.

For this analysis, a proton beam with an intensity of $7.5 \times 10^7$ particles per second and a kinetic energy of T = 4.5 GeV was incident on a $\sim$5 cm thick liquid-hydrogen target, resulting in a collection of $1.5 \times 10^9$ recorded events with HADES \cite{B}. This study focuses on the analysis of the exclusive process $p+p \rightarrow \Lambda + K^0_S + p + \pi^+$.

\newpage
\section{Analysis Techniques}

To ensure high-quality event selection, only events containing start time information and a reconstructed global event vertex were considered. The momenta of most final-state particles were reconstructed using the magnetic field and drift chamber information of HADES. For particles recorded in the forward detector, the momenta were calculated using time-of-flight (ToF), flight path measurements, and assuming the particle to be a proton. Momentum conservation was imposed to optimize the purity of the event selection. Specifically, the sum of the three ($x, y, z$) momentum components of two negative and four positive tracks was required to be within approximately 200 MeV/\textit{c} of the initial three-momentum of the beam. Additionally, the reconstructed primary vertex was required to be located in the $z$-direction (beam axis) downstream of the target origin to further improve event selection.

Particle identification (PID) was performed using time-of-flight information. Since negatively charged tracks were predominantly pions, a minimum velocity requirement of $\beta > 0.4$ was applied (with $\beta = v/c$). For positively charged tracks, which were primarily pions and protons, a relative time-of-flight method was employed to distinguish between species. Conventional PID techniques based on energy loss or timing information relative to the trigger, combined with momentum measurements, were insufficient to identify high-momentum proton tracks in the reaction of interest.

The time-of-flight value was measured by the Multiplicity and Electron Trigger Array (META) detector in HADES. The event start time, $t_{start}$, was determined through the innerTOF (iTof), and the stop time, $t^{stop}_i$, through the TOF or RPC detectors. The expected time-of-flight for a mass hypothesis was determined with the measured momentum and the distance traveled by each candidate particle. The $\pi^-$ track was selected as the reference candidate since it was the only negatively charged final-state particle after applying a veto selection from the RICH detector to remove the electrons. The event time offset was then determined by subtracting the expected from the measured time-of-flight of the $\pi^-$ tracks and then applied as a correction for all other tracks in the event.

This correction significantly improved the time resolution with respect to the iTof and, hence, the separation power between pions and proton tracks.
For the particle identification of the positively charged tracks, we studied the difference between the corrected time-of-flight measurement and the expected (calculated) time-of-flight based on the mass hypothesis and reconstructed momentum, referred to as $\Delta T$. The difference, $\Delta T$, of all charged-particle combinations for each event was obtained, assuming a mass hypothesis of two positively-charged pions and two proton candidates present in the channel of study. The combination that gave the smallest $\Delta T$ sum was used as the joint particle identification hypothesis for the event.

\begin{figure}[h]
\centering
\includegraphics[width=0.65\textwidth]{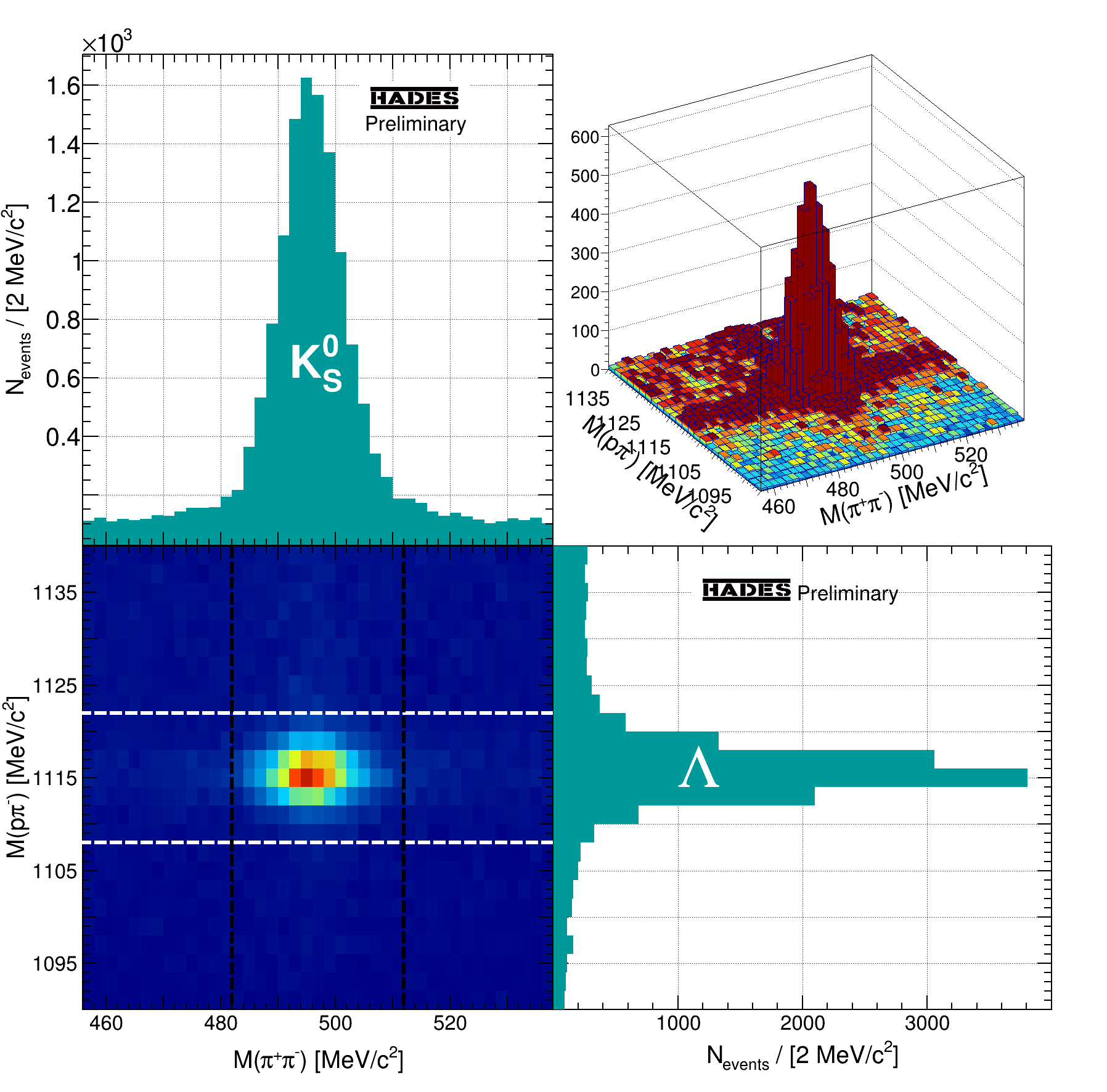}
\caption{Reconstructed invariant mass distribution of $M(p \pi^-)$ against $M(\pi^- \pi^+)$ along with their projections. Events presented in the $\Lambda$ and $K^0_S$ candidate projections are selected within the $M(\pi^- \pi^+)$ (white) and $M(p \pi^-)$ (black) band regions respectively.}
\label{fig:lambda_ks}
\end{figure}

Intermediate kaons and hyperons were reconstructed via decay channels, $K^0_S \rightarrow \pi^+ + \pi^-$ and $\Lambda \rightarrow p + \pi^-$. To enhance the selection of long-lived hyperons, a kinematic fitter based on vertex constraints was applied. Under the exclusive condition of detecting a primary pion and a proton, approximately 15,000 candidate events with $\Lambda$ and $K^0_S$ in coincidence were reconstructed as shown in Fig. \ref{fig:lambda_ks}. A similar selection was achieved by estimating the distance of closest approach (DCA) between the daughters of neutral candidates.

We subtracted uncorrelated background events with a sideband analysis on $\Lambda K^0_S$ and inspected the correlation between the invariant masses of $M^2(\Lambda K^0_S)$ against $M^2(p \pi^+)$ to identify contributions from intermediate resonances such as $\Delta^{++} \rightarrow p + \pi^+$. Since $\pi^+$ can also be combined with the other two particles that contain strangeness, we explored the possible contributions of their respective intermediate resonances, namely $K^* \rightarrow \pi^+ + K^0_S$ and $\Sigma^* \rightarrow \pi^+ + \Lambda$. 

We generated phase space distributed Monte Carlo data of the $p+p \rightarrow \Lambda + K^0_S + p + \pi^+$, $p+p \rightarrow \Lambda + K^0_S + \Delta^{++} (\rightarrow p \pi^+)$, $p+p \rightarrow \Sigma^* (\rightarrow \Lambda\pi^+) + K^0_S + p$, and $p+p \rightarrow \Lambda + K^* (\rightarrow K^0_S\pi^+) + p$ reactions using Pluto \cite{C}. The detector response was modeled with Geant3 \cite{D} and the resulting data were analyzed using the same conditions as applied to the experimental data. 

  
\section{Results}

\begin{figure}[h]
\centering
\begin{minipage}{0.45\textwidth}
    \centering
    \includegraphics[width=\textwidth]{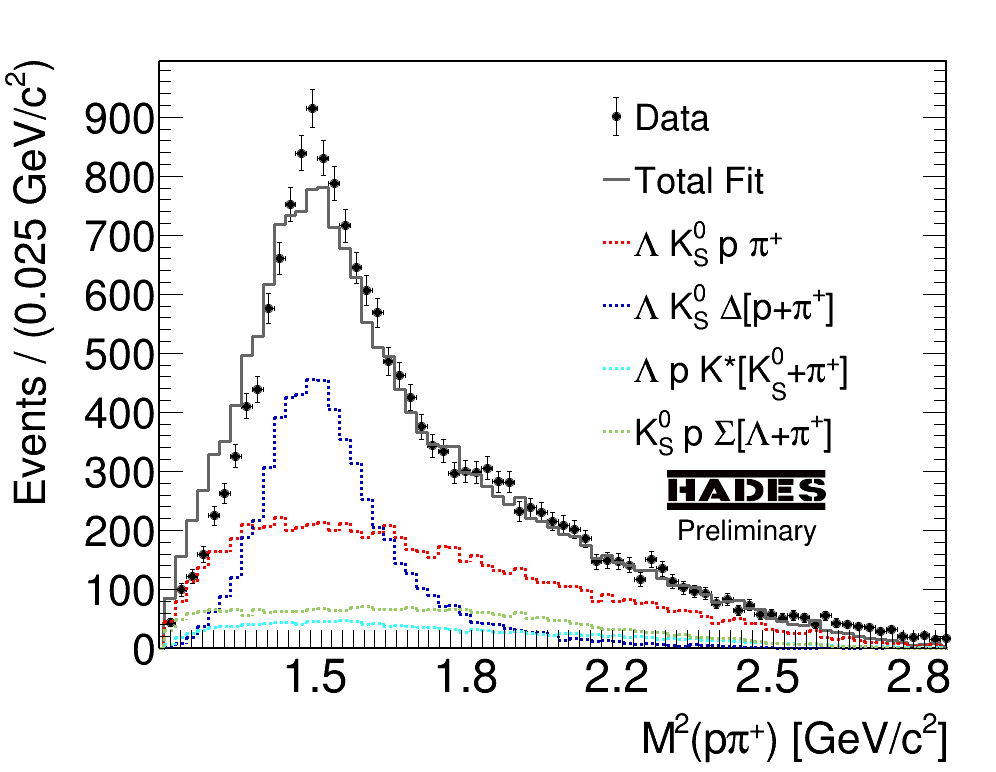}
    \subcaption{}
    \label{fig:Q1Q2}
\end{minipage}
\hfill
\begin{minipage}{0.45\textwidth}
    \centering
    \includegraphics[width=\textwidth]{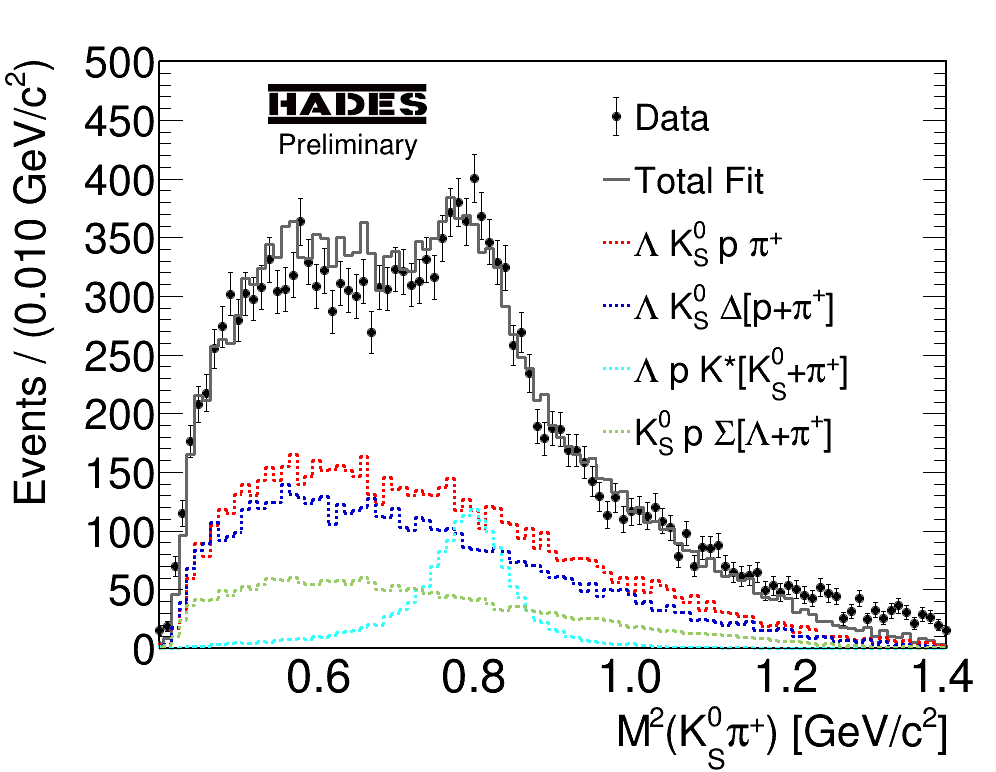}
    \subcaption{}
    \label{fig:Q3Q4}
\end{minipage}
\hfill
\begin{minipage}{0.45\textwidth}
    \centering
    \includegraphics[width=\textwidth]{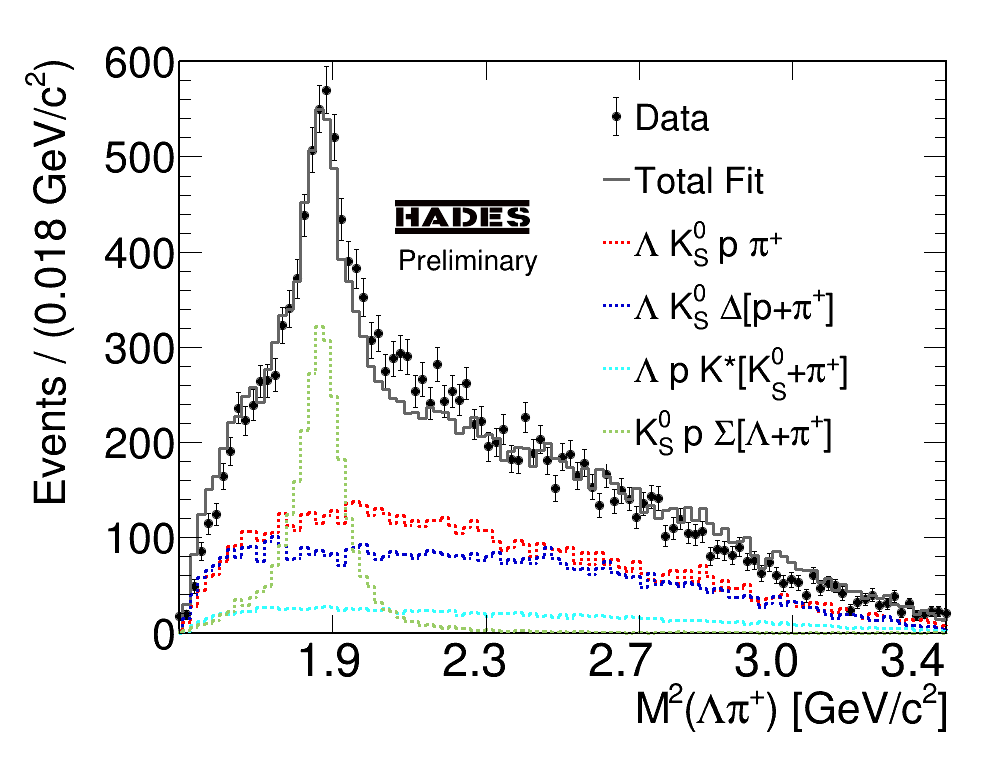}
    \subcaption{}
    \label{fig:Q5Q6}
\end{minipage}
\caption{Reconstructed distributions of $M^2(p \pi^+)$ (\subref{fig:Q1Q2}), $M^2(\pi^+ K^0_S)$ (\subref{fig:Q3Q4}), and $M^2(\pi^+ \Lambda)$ (\subref{fig:Q5Q6}), along with their relative contributions from the underlying intermediate 3-body and 4-body phase space contributions. The sideband-corrected data are shown as black markers.}

\label{fig:resonances}
\end{figure}

A comparison between signal Monte Carlo events and experimental data is presented in Fig.~\ref{fig:resonances}. A binned, simultaneous maximum likelihood fit is performed across three invariant mass regions by implementing a simplified model of the four contributions to the exit channel $p+p \rightarrow \Lambda + K^0_S + p + \pi^+$. The fit function is the incoherent sum of the events generated by the 4-body phase-space spectrum (red) and the three 3-body phase-space spectra with intermediate resonances---$\Delta^{++}$ (blue), $K^{*}$ (cyan) and $\Sigma^{*}$ (green)---whose relative contributions are treated as free parameters. The resulting fit (gray) yields a reduced $\chi^{2}/ndf = 4.4$.

The data and fit results shown in Fig.~\ref{fig:resonances} demonstrate clearly the presence of three intermediate resonances, namely $\Delta^{++} (\rightarrow p + \pi^+)$, $K^* (\rightarrow \pi^+ + K^0_S)$, and $\Sigma^* (\rightarrow \pi^+ + \Lambda)$, compatible with their expected masses as reported by the Particle Data Group \cite{E}. Although this incoherent fit provides a reasonable description of the data, an improved fit is required for a more quantitative understanding of the underlying dynamics.



\section{Conclusion}

In conclusion, this study provides clear evidence of intermediate resonance contributions in the reaction $p+p \rightarrow \Lambda + K^0_S + p + \pi^+$ at T = 4.5 GeV. We observed the decay of $\Delta^{++} \rightarrow p\pi^+$ and $\Sigma^* \rightarrow (\Lambda\pi^+)$ enabling us to study the role of intermediate baryon resonances via the channel $pp\rightarrow \Lambda K^0_S \Delta^{++}$ and $pp \rightarrow \Sigma^* K^0_S p$. This serves as a complementary probe to earlier studies based upon the $pp\rightarrow \Lambda K^+ p$ and $pp\rightarrow \Sigma^0 K^+ p$ channels \cite{A}. Additionally, the observation of the vector-meson resonance $K^* \rightarrow (\pi^+ K^0_S)$ opens the possibility to extract the polarization of $K^*$ via the spin density matrix elements \cite{F}. This study will be followed up by a more detailed and quantitative analysis using a partial wave amplitude analysis.


\begin{thebibliography}{99}

\bibitem{A}
R. Abou Yassine \textit{et al}., Eur. Phys. J. A \textbf{60}, 18 (2024).



\bibitem{B}
G. Agakichiev \textit{et al}., Eur. Phys. J. A \textbf{41}, 243–277 (2009).


\bibitem{C}
    I. Fröhlich \textit{et al}.,
   J. Phys Conf. Ser. \textbf{219} 032039 (2010).
   
\bibitem{D}
Y. Zheng-Yun \textit{et al}., 
Chinese Phys. C \textbf{32} 572 (2008).


\bibitem{E}
P.A. Zyla \textit{et al}., (Particle Data Group), Prog. Theor. Exp. Phys. \textbf{2020}, 083C01 (2020).

\bibitem{F}
J. F. Donoghue, Phys. Rev. D \textbf{17}, 2922 (1978).

\end{thebibliography}
\end{document}